\documentclass{article}
\usepackage{graphicx} 

\usepackage{xcolor}
\usepackage{hyperref}
\usepackage{titletoc}

\definecolor{mygreen}{RGB}{0,128,0}

\usepackage{array} 
\usepackage{booktabs} 
\usepackage{graphicx} 
\usepackage{float}

\usepackage[square,numbers]{natbib}

\usepackage[margin=1in]{geometry}

\begin{document}

\begin{center}
    {\Large \textbf{Lensed hot stars with HST in the 2030s} \par}
    \vspace{0.2cm}
    J.M. Diego$^1$\footnote{jdiego@ifca.es} \\
    \vspace{0.5cm}
    $^1$ Instituto de F\'isica de Cantabria (CSIC-UC). Avda. Los Castros s/n. 39005 Santander, Spain. \\
     \vspace{0.5cm}

    \today
     \vspace{0.5cm}
\end{center}

\begin{abstract}
In the late 20th century, the Hubble Space Telescope (HST) revolutionized astronomy, showing the Universe with a detail never seen before in the ultraviolet (UV),  optical and infrared (IR) bands. In the early 2020s, the James Webb Space Telescope started a similar revolution in the IR. The launch of Roman in late 2026 challenges the reign of HST in the optical band, but even after Roman's launch, HST will remain as the only telescope capable of high-quality imaging in the UV band. In the optical bands, HST provides superior resolution than Roman for point sources. Although equipped with more sensitive MOS, Roman's sensors have a pixel size about 3 times larger than HST's CCDs, hence undersampling the point-spread-function, and resulting in a worse spatial resolution. The UV-capable and higher-resolution of HST in the UV and optical band, makes HST the best instrument for specific science cases. This paper responds to the "Building a Roadmap for Hubble science into the 2030s" call and focuses on science with lensed hot stars at $z>0.5$ in the UV and optical bands,  exploiting the features that makes HST the best instrument in the UV/optical until the launch of the Habitable World Observatory in the 2040s.
\end{abstract}







\section{Introduction}
With the Hubble Space Telescope (HST) quickly approaching the point of orbital atmospheric re-entry in 2033, STScI recently made a call for community input to help define a new roadmap for the programmatic, operational, and scientific future of Hubble into the next decade, hence supporting a possible boost to extend HST's lifetime into the 2040s. This white paper is part of a series of papers that answer this call, by describing a specific science case where HST will still remain the best telescope in the 2030s. This white paper focuses on a particular type of transient phenomena in the UV and optical bands; {\it Lensed hot stars at cosmological distances} ($z>0.5$). \\

Observing luminous hot stars at $z>0.5$ is possible thanks to the boost provided by gravitational lensing near the critical curves of galaxy clusters, where the magnification factor can reach thousands (equivalent to 8--10 mags). These stars appear as transients due to the short duration of microlensing events (days to weeks) that momentarily boost their magnification beyond the detection limit. Hot blue supergiants are best studied with HST since their emission peaks in the UV and optical bands. Their smaller radii, compared with the much larger but cooler red supergiant stars, allows blue supergiants to be magnified more (almost two magnitudes more) than their red relatives. This follows from the dependency of the maximum magnification, $\mu_{\rm max}$, with, $R$, the star radius, $\mu_{\rm max}\propto (\sqrt{R})^{-1}$. Compared with Roman, and relevant for point source science, HST's has a greater energy concentration power due to its better spatial sampling (0".04 in HST's WFC3 UVIS/Opt vs 0."11 in Roman's WFI). The point source nature of lensed hot stars and their strong UV emission demands the highest possible angular resolution at the shortest possible wavelengths. \\

These stars are key to understand the star formation history of galaxies beyond $z>0.5$, which is strongly constrained by the presence of massive hot stars. The compact nature of these stars can also be used as pencil beams mapping the small scale substructure in the lens plane, making it possible to constrain certain dark matter models. {\bf HST will dominate this field until the launch of the Habitable World Observatory (HWO) in the 2040s}.

\section{Lensed Blue Supergiant stars}
HST has played a pioneering role in this field. The first discovery of a lensed star beyond our local universe, Icarus ($z=1.49$), took place during an HST  monitoring campaign of the lensed SN Refsdal behind the galaxy cluster MACS1149 \cite{Kelly2018}.
Icarus, is the very first example of of an individual star being strongly lensed  beyond our local universe ($D_l<20$ Mpc or $z<0.005$) and pushing the record distance for the farthest star by over two orders of magnitude ($D_l>2000$ Mpc or $z>0.5$). The discovery of Icarus represents the birth of a new field in astronomy, the study of individual stars at redshifts $z>0.5$. After the discovery of Icarus, many more stars were discovered with HST \cite{Rodney2018,Chen2019,Kaurov2019,Diego2022_Godzilla,Kelly2022,Chen2022b}, culminating with the discovery of Earendel, the farthest star candidate ever discovered at $z\approx 6$  \cite{Welch2022}. These stars have allowed for a variety of new studies. \\

\begin{figure} 
   \includegraphics[width=16.0cm]{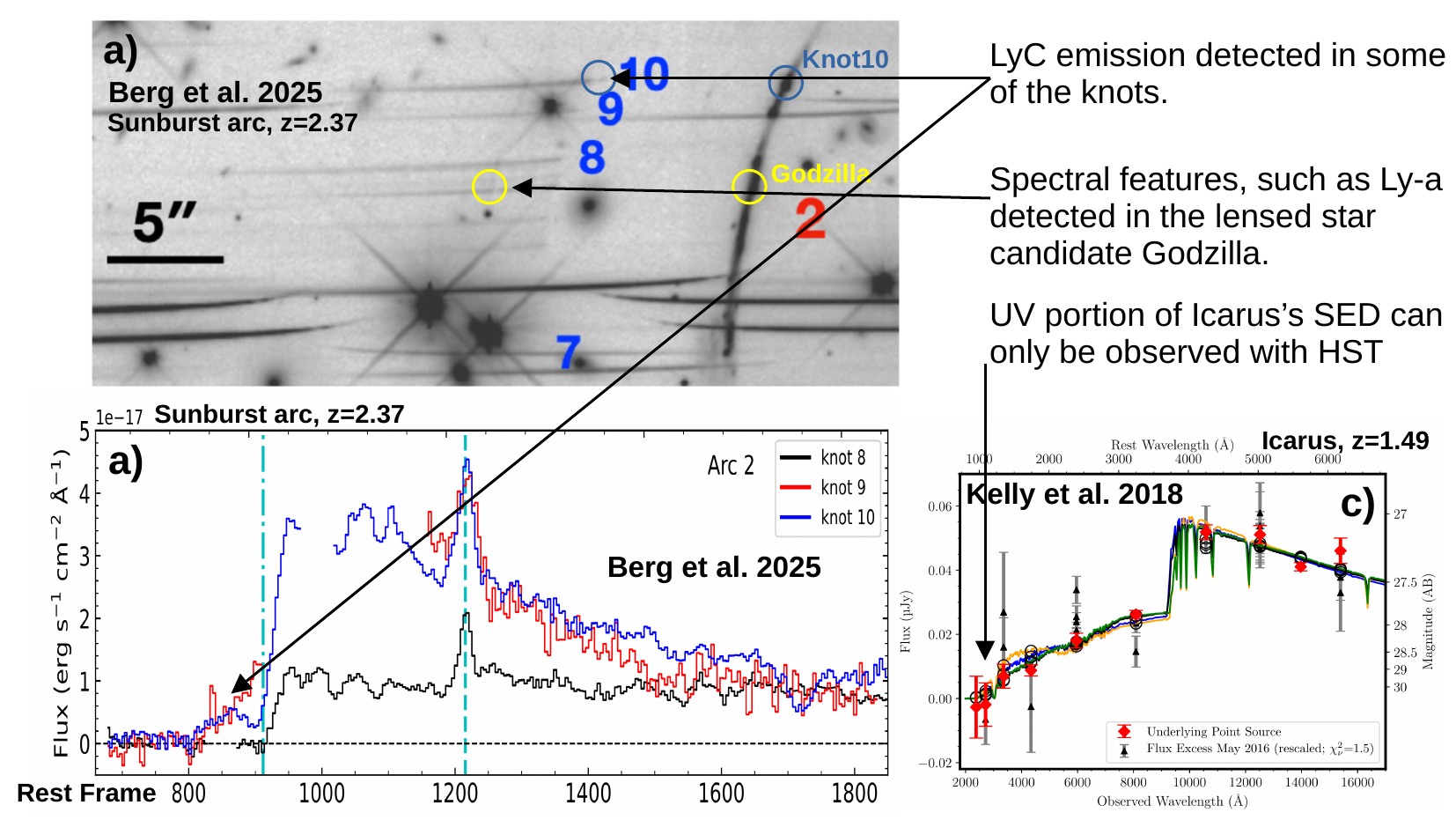}
      \caption{HST/WFC3 UVIS G280 spectra. 
HST spectra shows leaking LyC photons at redshift z=2.37 from some of the knots in the Sunburst arc. Ly-a is also detected in all knots.  Godzilla does not show LyC emission, but the Ly-a peak is clearly detected. Panel c) shows the SED of Icarus as measured by HST. The colored curves are models for stars with temperatures $\approx 12000$ K. {\bf HST is the only instrument capable of making this kind of measurement down to the UV from space.} }
         \label{Fig_1}
\end{figure}

{\bf Galaxy evolution and IMF:} The most luminous lensed star candidate ever observed, Godzilla, was also discovered thanks to HST and extreme magnification \cite{Diego2022_Godzilla}. 
HST's UV capabilities are key to interpret this and other stars. As shown in Figure~\ref{Fig_1}, hyperluminous stars such as Godzilla show intense emission in the UV part of the spectrum which is best studied by HST. In particular, the UVIS GRISM spectra has been used to detect Ly-$\alpha$ emission in Godzilla, and other interesting spectral features in compact regions near Godzilla, including highly ionizing photons from the Lyman continuum  portion of the spectra \cite{RiveraThorsen2019}. \\

Thanks to the high-redshift, large magnification (thousands) , and extraordinary intrinsic luminosity of lensed star candidates such as Godzilla ($z=2.37$), they can still have part of their UV spectra observed from the ground. However, most lensed stars discovered so far have more modest redshifts between $z\approx 1$ and $z\approx 2$. For these stars, features such as the Ly-$\alpha$ line or Lyman break are absorbed by the ozone layer and cannot be observed from the ground. HST is the only instrument that remains with the spectral coverage, spatial resolution, and sensitivity to make this type of observation. \\

\begin{figure}[tbh]
    \begin{minipage}{0.56\textwidth}
          \includegraphics[width=9.5cm]{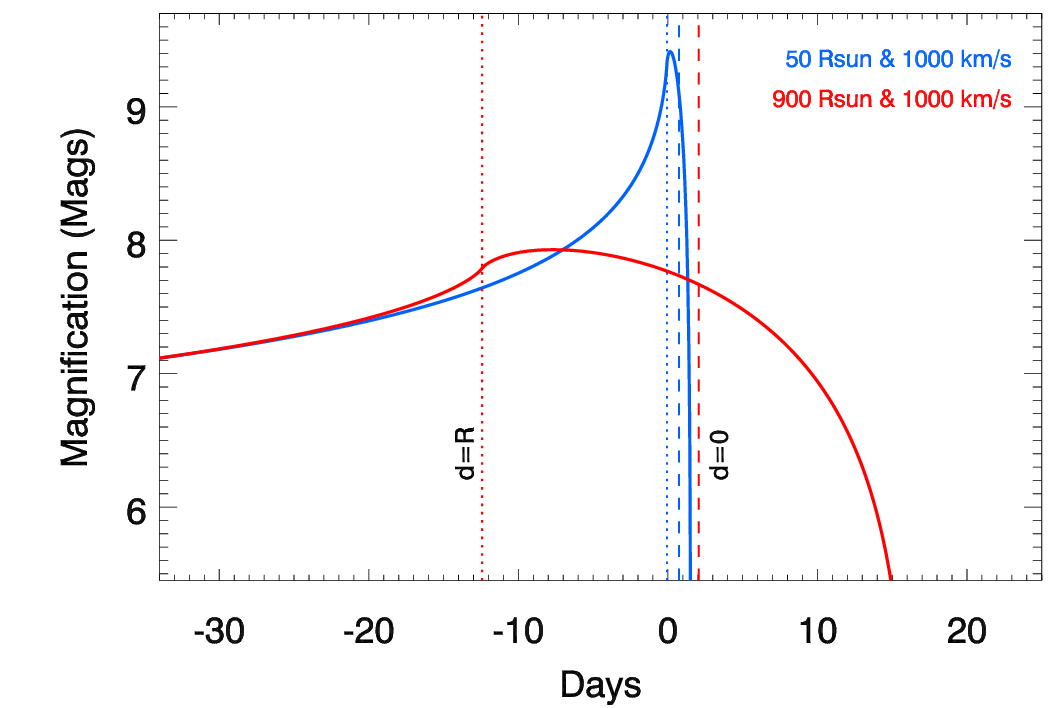}  
    \end{minipage}%
    \hspace{\fill}%
    \begin{minipage}{0.4\linewidth}
        \caption{Magnification (expressed in magnitudes) of two stars moving with  velocity 1000 km s$^{-1}$a toward a caustic (blue $R=50\ {\rm R}_{\odot}$, red $R=900\ {\rm R}_{\odot}$). Owing to the much smaller size of hot blue supergiants, they can be magnified more and be observed at farther distances. The vertical dotted and dashed lines show the moments where the star center is at one star radius (d=R) and in the middle of the caustic (d=0), respectively. {\bf HST's better ability to detect blue supergiants allows us to see these stars farther away than  red supergiant stars.}}
        \label{Fig_2}
    \end{minipage}
\end{figure}

So far only a handful lensed galaxies below $z=1$ have revealed lensed stars in their interiors. Most notably, the Dragon arc, at $z=0.725$ (and the lowest-redshift caustic-crossing galaxy known so far),  has revealed tens of lensed stars \cite{Fudamoto25,Palencia2026}. These low redshift galaxies are of particular interest because their smaller distance modulus always result in a larger number of detections of microlensing events.  \\

In the next decade, the number of known low redshift ($z<1$) caustic crossing galaxies will increase exponentially thanks to the new deep large scale surveys; Euclid, LSST, and Roman. These caustic crossing galaxies will be followed up with telescopes such as JWST but also HST, revealing their population of blue and red supergiant stars. Although many of these stars will still be too faint for the UVIS GRISM, the photometry will help determine their temperatures. The power of HST to measure the temperature of lensed stars was beautifully demonstrated with Icarus, as shown in panel c) of Figure~\ref{Fig_1}. HST's UV coverage is essential to correctly classify the hottest (O and B) lensed stars. \\

While JWST is very powerful at detecting red supergiant stars, HST is better suited to study the hot luminous stars. Detecting both types of stars is important to better understand the evolutionary state of the galaxy hosting them. The blue-to-red ratio of luminous stars complements studies of the IMF and star formation history based on SED fitting \cite{Li2025}. \\

{\bf Dark matter:} These stars, as interesting as they are for stellar astrophysics and galaxy evolution, can be also used to study dark matter, specially on the smallest scales. As the observer, lens, and distant star planes moves with respect to each other, small perturbations in the lens plane move in front of the line of sight from the observer to the distant star, changing the magnification of the star, and hence the observed flux. The light curves from these stars can then be used to map the distribution of dark matter and to set limits on some models. This new type of constraint on dark matter models was first attempted with the first light curves of Icarus obtained with HST data \cite{Oguri2018}. Similar studies to exploit these stars as probes of dark matter models have been suggested \cite{Venumadhav2017,Diego2018,Dai2020,Williams2024,Diego2024b,Ji2025}. \\


To this day, and despite the success of the James Webb Telescope (JWST) at detecting individual stars beyond $z=1$, HST's Earendel remains as the record holder for the most distant star.  Part of the reason behind this long standing record is the fact that HST is much more efficient than JWST at detecting the hot blue supergiants, as Earendel. JWST is very powerful at detecting red supergiants, with over 100 such stars detected by JWST so far \cite{Meena2023a,Meena2023b,Diego2023a,Diego2023c,Yan2023,Furtak2024,Fudamoto25,Williams2026a,Williams2026b,Diego2026,Palencia2026,Furtak2026}. Most of the stars detected by JWST are cool red supergiants, better seen with JWST's IR-capable sensors, but the task of detecting the hottest and most luminous blue supergiants belongs to HST. Dedicated HST programs such as FLASHLIGHTS have proven HST's capacity to detect blue supergiants \cite{Kelly2022}.  Similarly to red supergiants, blue supergiants stars are among the most luminous in the universe, but contrary to red supergiants they are much smaller with radii in the range of tens or solar radii. Meanwhile red supergiants have radii in the hundreds of solar radii, and up to $R_{\rm max}\sim 1500\, {\rm R}_{\odot}$, making them as luminous as the hot blue supergiants. This difference in radius explains why Earendel is still holding the record for the most distant star. The maximum magnification that a star can reach scales as $(\sqrt{R} )^{-1}$ \cite{MiraldaEscude1991}. Hence a $R=30\ {\rm R}_{\odot}$ blue supergiant observed during maximum magnification can be $\approx 2$ magnitudes brighter than a red supergiant during maximum magnification, similar intrinsic luminosity (before magnification), but with a much larger radius, $R_{\rm max}\approx1500\ {\rm R}_{\odot}$. Blue supergiants, and other hot luminous stars (including Wolf-Rayet stars), are extraordinarily precise clocks measuring recent star formation, and hence their observation in distant galaxies provide unique insight into the physical processes in those galaxies, that could not be obtained otherwise. \\

A demonstration of the difference in maximum magnification for red and blue supergiants is shown in Fig.~\ref{Fig_2}. We have assumed the same typical velocity relative to the caustic for both stars, $v=1000$ km s$^{-1}$. The magnification, $\mu$, scales with the distance, $d$, to the caustic as $\mu\propto1/\sqrt{d}$. The blue supergiant can get closer to the caustic reaching maximum magnification at $d \approx R_{\rm blue}$, while the red supergiant reaches maximum magnification farther away from the caustic at $d \approx R_{\rm red}>>R_{\rm blue}$, and hence with a smaller magnification. The change in flux in the blue supergiant changes very rapidly when its crossing the caustic, and in a very predictable way, making it easier to identify as a genuine caustic crossing event. The comparison shown in the figure is even more extreme when comparing even smaller blue supergiants with the largest red supergiants that can reach radii $R_{\rm max}\approx 1500\ {\rm R}_{\odot}$. \\

Blue supergiants at high redshift are very valuable for future studies that can be complemented with other telescopes (JWST studies red supergiant stars while Roman will find young and blue caustic crossing galaxies that can be later  followed with HST). The redshifted peak emission of hot stars formed in the cosmic noon can best be studied with HST. Detecting and studying hot stars during this important phase in the history of the universe can shed light into studies made with telescopes such as JWST. Massive hot stars are expected to play an important role during cosmic noon. As discussed earlier, given their smaller size, hot blue stars can be magnified more than red supergiants, thus allowing us to detect them farther away or detecting stars that are intrinsically fainter than red supergiant stars. Moreover, it has been well established that red supergiants can not exceed luminosities of $\approx 6\times10^5 {\rm L}_{\odot}$, known as the Humphreys-Davidson limit \cite{HD1979}. On the contrary, blue supergiants can be brighter than $\approx 10^6 {\rm L}_{\odot}$ making them easier to find at high redshifts. \\

Even at magnification factors of thousands, these lensed stars remain well below the resolution power of any telescope and appear as point sources. Detecting them requires the best possible spatial resolution. In the blue and UV bands, where these stars are brightest, no other telescope can compete with HST. Even its sibling telescope, Roman. won't be able to compete with HST. At a pixel scale with $0".11$ sampling (or twice worse than HST), Roman will undersample the point-spread-function in the bluest bands and achieve effective resolutions that are worse than HST's in the bluest filters. 
More importantly is the spectral coverage in the UV provided by HST. Hot blue stars at $1<z<2$ have their peak emission below 4000 Angstroms, beyond the reach of Roman. Hot blue stars in this redshift range have been detected by HST and offer unique opportunities for science. High cadence monitoring of these stars can be used to constrain models of dark matter in unique ways. For example, axion minihalos can leave a modulation in the light curve of a caustic crossing star that are much easier to measure with the compact blue supergiants than with the large red supergiants \cite{Dai2020}. \\

\begin{figure}[tbh]
    \begin{minipage}{0.43\linewidth}
        \caption{Figure taken from \cite{Diego2026}. The small panel on the right side shows a realization of the magnification pattern (and millicaustics) produced by a model of wave dark matter ($m_a\sim 10^{-22}$ eV, $\lambda_{\rm dB}=10$ pc) near a galaxy cluster critical curve. Microlenses will perturb this pattern even further on microarcsecond scales (not shown, microcaustics). The probability of magnification is shown in the main panel. Hot blue stars moving in this field of wave dark matter millicaustics and microcaustics will sample both the microlens and dark matter distributions. {\bf HST's detected hot blue stars are better than red cool stars to map the small substructures}. }
        \label{Fig_3}
    \end{minipage}
    \begin{minipage}{0.56\textwidth}
          \includegraphics[width=9.5cm]{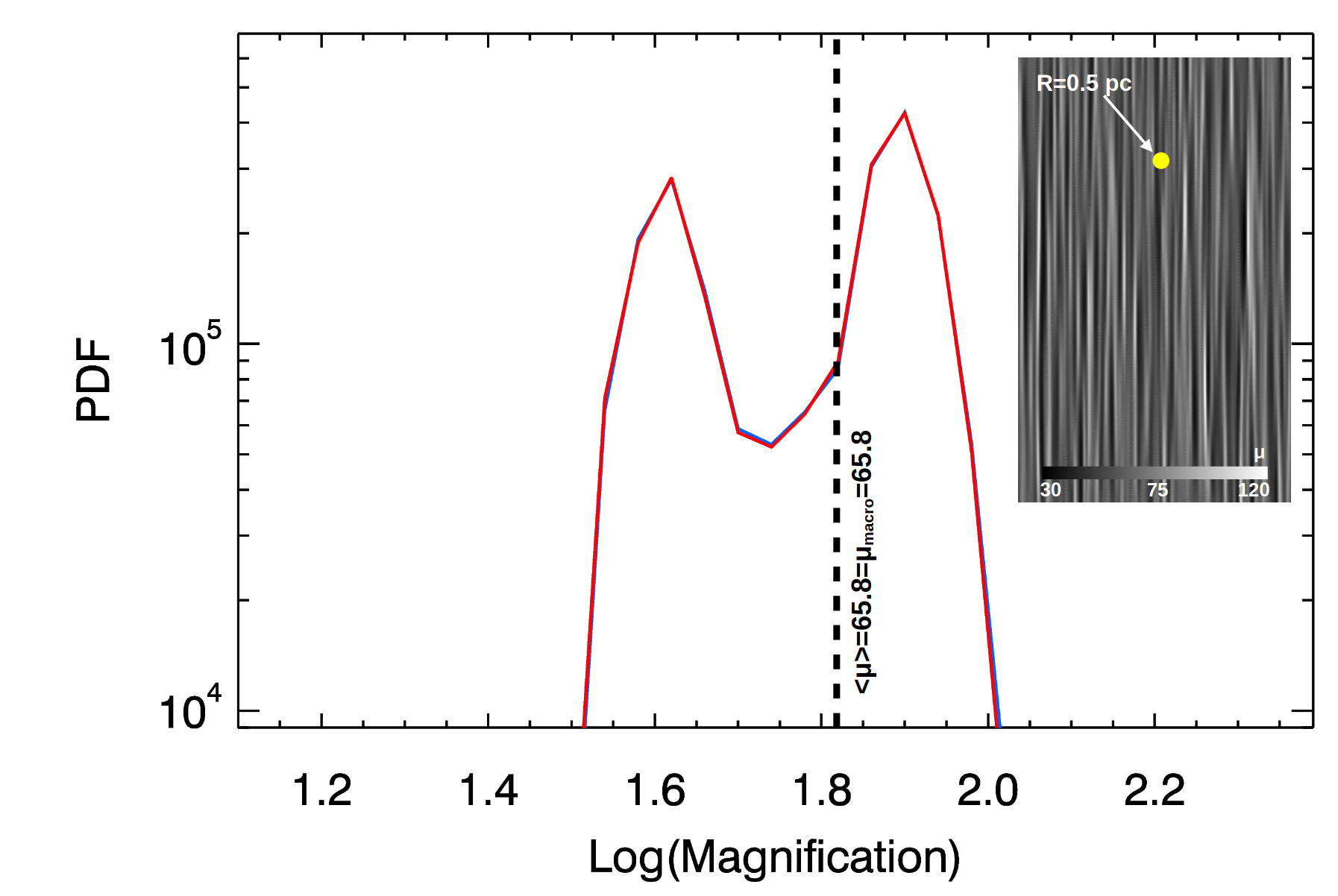}  
    \end{minipage}%
\end{figure}
The small radii of blue supergiants allow us also to detect very small scale perturbations, on mass scales comparable to planets, but at redshifts $z>0.5$ and during the a microcaustic crossing event \cite{Diego2018}. Instead of planets, one can consider dark matter candidates in the same mass regime, such as primordial black holes \cite{Oguri2018}. Such small masses are very difficult to probe with red supergiants using JWST, but are much easier to detect with the smaller hot luminous stars using HST. 
The higher sensitivity of compact blue luminous stars to small scale perturbations provides unique opportunities to test other models of dark matter. For instance, wave dark matter is often discussed in the context of very small axion masses, $m_a\sim 10^{-22}$ eV. At these masses, wave dark matter produces millicaustics near the critical curves of galaxy clusters (see Figure~\ref{Fig_3}). The scale of these millicaustics is $\sim 1$ parsec in one dimension and many parsecs in the perpendicular direction. But axion masses as small as $m_a\sim 10^{-22}$ eV are in tension with observations of the Ly-$\alpha$ forest and dwarf galaxies (plus other observations), with current limits suggesting that if dark matter is a scalar field, the associated particle must be heavier than  $m_a\sim 10^{-19}$ eV. For such heavier axion masses, the associated de Broglie wavelength in a galaxy cluster is very small, $\lambda_{\rm dB}\sim 0.01$ pc, and the projected wave DM fluctuations have dispersion $\sigma_{\rm dB}\sim 0.1\, {\rm M}_{\odot}$. The caustic size of the wave dark matter fluctuations reduces to $0.001$ pc, that is smaller than the typical size of microcaustics from stars. These perturbations in the wave dark matter field are very small and difficult to detect. However, they get amplified near the critical curves of clusters. Lensed hot blue stars, owing to their smaller size, are the best way to probe these smaller substructures. Microlenses near the cluster critical curves will overwhelm this signal in general, but near their associated microcaustics, the small fluctuations produced by wave dark matter will be also amplified by the microlenses. Wave dark matter fluctuations are ubiquitous and should be observed in all microcaustic crossings. The small wave dark matter fluctuations will produce their own network of microcaustics with scales in the source plane of a fraction of a microarcsecond, which will modulate the more intense effect produced by the more massive, and concentrated, microlenses (stars near the galaxy cluster critical curve). Microcaustics from microlenses will also be ubiquitous very close to the galaxy cluster caustic. Mapping the perturbations at the microarcsecond scale is possible with  repeated observations made with HST. {\bf One to three years monitoring of a microcaustic event, with a cadence of one HST observation every month, or every two months, may be sufficient to clearly establish the existence of a modulation in the magnification from a scalar field (wave dark matter).} Even better,  monitoring low-z caustic crossing galaxies such as the Dragon arc (with O(100) microlensing event candidates already reported in the literature) can increase the statistics (and hence significance of the detection of wave dark matter and its associated axion mass) since multiple microcaustics events can be monitored all at once. {\bf The higher spatial  resolution in the source plane, and larger magnification factors achieved by compact hot blue stars (when compared to large red supergiants) makes HST the ideal instrument to prove wave dark matter models for the still unconstrained regime with axion masses greater than $m_a\sim 10^{-19}$ eV}.\\

HST will remain the champion telescope in the 2030s for studying highly magnified hot blue stars at cosmological distances. These stars give us a different perspective into the evolution of galaxies at these redshifts, traced by their most massive and luminous young stars. But they also serve as probes of the small scale structure in the lens plane, with great potential to probe dark matter models that can only be tested in extreme scenarios (for instance via superradiance of supermassive black holes, \cite{Stott2018}). The discovery of what dark matter really is may still be far away in time, but the key clues about its true nature may be just around the corner. We just need to use the best magnifying glass to find those clues. What a shame would be to let it burn... 


\bibliographystyle{apj}
\bibliography{Biblio} 

\end{document}